\documentclass{elsart}   

\usepackage{graphicx}

\begin{document}

\begin{frontmatter}
\title{Inverse Statistics in Economics : The gain-loss asymmetry}

\author{Mogens H. Jensen},
  \ead{mhjensen@nbi.dk}
\author{Anders Johansen\thanksref{anders-now}},
  \ead{anders.johansen@risoe.dk}
  \thanks[anders-now]{Present address: Ris{\o} National Laboratory,
           Wind Energy Department, P.O. 49, DK-4000 Roskilde, Denmark}
  and  
\author{Ingve Simonsen}
  \ead{ingves@nordita.dk}

\address{The Niels Bohr Institute and NORDITA,\\
     Blegdamsvej 17, DK-2100 Copenhagen {\O},\\  DENMARK }

 
\begin{abstract}
  Inverse statistics in economics is considered. We argue that the
  natural candidate for such statistics is the investment horizons
  distribution. This distribution of waiting times needed to achieve a
  predefined level of return is obtained from (often detrended)
  historic asset prices. Such a distribution typically goes through a
  maximum at a time called the {\em optimal investment horizon}, $\tau^*_\rho$,
  since this defines the most likely waiting time for obtaining a given 
  return $\rho$.  By considering equal positive and negative levels of 
  return, we report on a quantitative gain-loss asymmetry most pronounced 
  for short horizons. It is argued that this asymmetry reflects the market
  dynamics and we speculate over the origin of this asymmetry.
\end{abstract}


\end{frontmatter}


%
%
Financial time series have been recorded and studied for many decades.
With the appearance of the computer, this development has accelerated,
and today large amounts of financial data are recorded daily. These
data are used in the financial industry for statistical studies and
for benchmarking.  In particular, they can be used to measure the
performance of a financial instrument. Traditionally this has been
done by studying the distribution of
returns~\cite{Book:Bouchaud-2000,Book:Mantegna-2000,Book:Hull-2000}
calculated over a {\em fixed} time period $\Delta t$.  Such
distributions measure how much an initial investment, made at time
$t$, has gained or lost by the time $t+\Delta t$. Numerous empirical
studies have demonstrated that for not too large $\Delta t$'s, say
from a few seconds to weeks, the corresponding (return) distributions
are characterized by so-called fat
tails~\cite{Book:Bouchaud-2000,Book:Mantegna-2000,Book:Hull-2000,Mandelbrot-1967}.
This is to say that the probability for large price changes are much
larger then what is to be expected from Gaussian statistics, an
assumption typically made in theoretical and mathematical
finance~\cite{Book:Bouchaud-2000,Book:Mantegna-2000,Book:Hull-2000}.
However, as $\Delta t$ is increased even further, the distribution of
returns gradually converge to the Gaussian distribution.

In the context of economics, it was recently
suggested~\cite{Simonsen-EPJB-2002}, partly inspired by earlier work
in turbulence~\cite{Jensen-PRL-1999}, to alternatively study the
distribution of waiting times needed to reach a {\em fixed} level of
return. These waiting times, for reasons to be clarified in the
discussion below, were termed {\em investment horizons}, and the
corresponding distributions the {\em investment horizon
  distributions}. Furthermore, it was shown for positive levels of
return, that the distributions of investment horizons had a
well-defined maximum followed by a power-law tail scaling
like\footnote{Notice that this scaling behavior implies that the first
  (average investment horizon), and higher, moments of this
  distribution do not exist.}  $p(t)\sim t^{-3/2}$. The maximum of
this distribution signifies the {\em optimal investment horizon} for
an investor aiming for a given return.

\begin{figure}[t]
    \begin{center}
        \leavevmode
        \includegraphics*[width=0.8\columnwidth,height=0.43 \columnwidth]{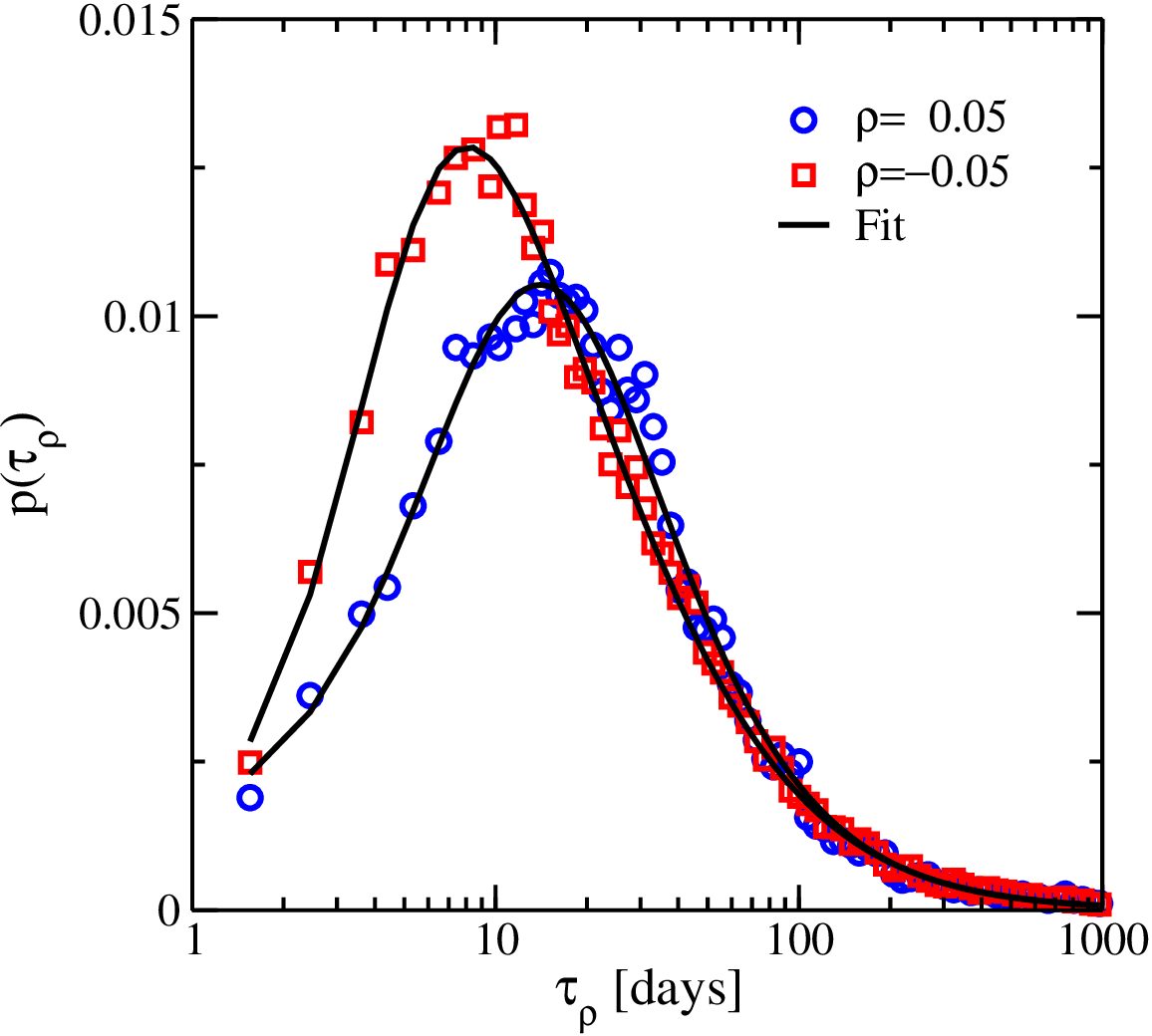} 
       \caption{The investment horizon distributions for the DJIA
         closing prices at a return level $\left|\rho\right|=0.05$.
         The open symbols correspond to the empirical distributions,
         while the solid lines represents the maximum likelihood fit
         of these distributions to the functional form given by
         Eq.~\protect(\ref{fit-func}).  The fitting parameters used to
         obtain these fits are for $\rho=0.05$: $\alpha=0.50$,
         $\beta=4.5$~days$^{1/2}$, $\nu=2.4$, and $t_0=11.2$~days; and
         for $\rho=-0.05$: $\alpha=0.50$, 
         $\beta=5.0$~days$^{1/2}$, $\nu=0.7$, and $t_0=0.6$~days.  
          }
        \label{Fig:DJIA_PDF}
    \end{center}
\end{figure}

%
%
In order to present the method, let us start by letting $S(t)$ denote
the asset price. Then the logarithmic return at time $t$, calculated
over a time interval $\Delta t$, is defined
as~\cite{Book:Bouchaud-2000,Book:Mantegna-2000,Book:Hull-2000}
\begin{eqnarray}
    \label{Return}
    r_{\Delta t}(t) &=& s(t+\Delta t)- s(t),
\end{eqnarray}
where $s(t) = \ln S(t)$. Hence the log-return is nothing but the
log-price change of the asset. We consider a situation where an
investor is aiming for a given return level denoted $\rho$, which may
be both positive (being ``long'' on the market) or negative (being
``short'' on the market).  If the investment is made at time $t$, then
the investment horizon is defined as the time $\tau_\rho(t)=\Delta t$
so that the inequality $r_{\Delta t}(t)\geq \rho$ when $\rho\geq 0$,
or $r_{\Delta t}(t)\leq \rho$ when $\rho<0$, is satisfied for the {\em
  first} time. The investment horizon distribution, $p\left(
  \tau_\rho\right)$, is then the distribution of investment horizons
$\tau_\rho$ (see Fig.~\ref{Fig:DJIA_PDF}) averaged over the data.

A classic assumption made in theoretical finance is that the asset
prices follow a geometrical Brownian motion, {\it i.e.} $s(t)=\ln
S(t)$ is just a Brownian motion. For a Brownian motion, the investment
horizon (first passage time) problem is known
analytically~\cite{Book:Karlin-1966,Rangaraja-PLA-2000}. It can be
shown that the investment horizon distribution is given by the
Gamma-distribution: $p(t) =
\left|a\right|\exp(-a^2/t)/(\sqrt{\pi}t^{3/2})$, where $a\propto
\rho$. Note, that in the limit of large (waiting) times, one recovers
the well-known first return probability $p(t) \sim t^{-3/2}$.  As the
empirical logarithmic stock price process is known not to be
Brownian~\cite{Book:Bouchaud-2000,Book:Mantegna-2000,Book:Hull-2000,Mandelbrot-1967},
we instead suggest to use a generalized (shifted) Gamma distribution
of the form:
\begin{eqnarray}
    \label{fit-func}
    p(t) &=&
    \frac{\nu}{\Gamma\left(\frac{\alpha}{\nu}\right)}\,
    \frac{\left|\beta\right|^{2\alpha}}{(t+t_0)^{\alpha+1} }
    \exp\left\{
          -\left(\frac{\beta^2}{t+t_0}\right)^{\nu} 
        \,\right\},
\end{eqnarray} 
%
%
as a basis for fitting the empirical investment horizon distributions.
It will be seen below, that this form parametrize the empirically data
excellently.  Note, that the distribution, Eq.~(\ref{fit-func}),
reduces to the Gamma-distribution (given above) in the limit of
$\alpha=1/2$, $\beta=a$, $\nu=1$, and $t_0=0$.  Furthermore, the
maximum of this distribution, {\it i.e.}  the {\em optimal investment
  horizon}, is located at $\tau^*_\rho =
\beta^2(\nu/(\alpha+1))^{1/\nu} -t_0$ for a given level of return
$\rho$.
If the underlying asset price process is geometric Brownian, then one would 
have $\tau^*_\rho \sim \rho^2$ for {\em all} values of $\rho$. We will later 
see that this is far from what is observed empirically.

\begin{figure}[t]
    \begin{center}
        \leavevmode
         \includegraphics*[width=0.8\columnwidth,height=0.43\columnwidth]{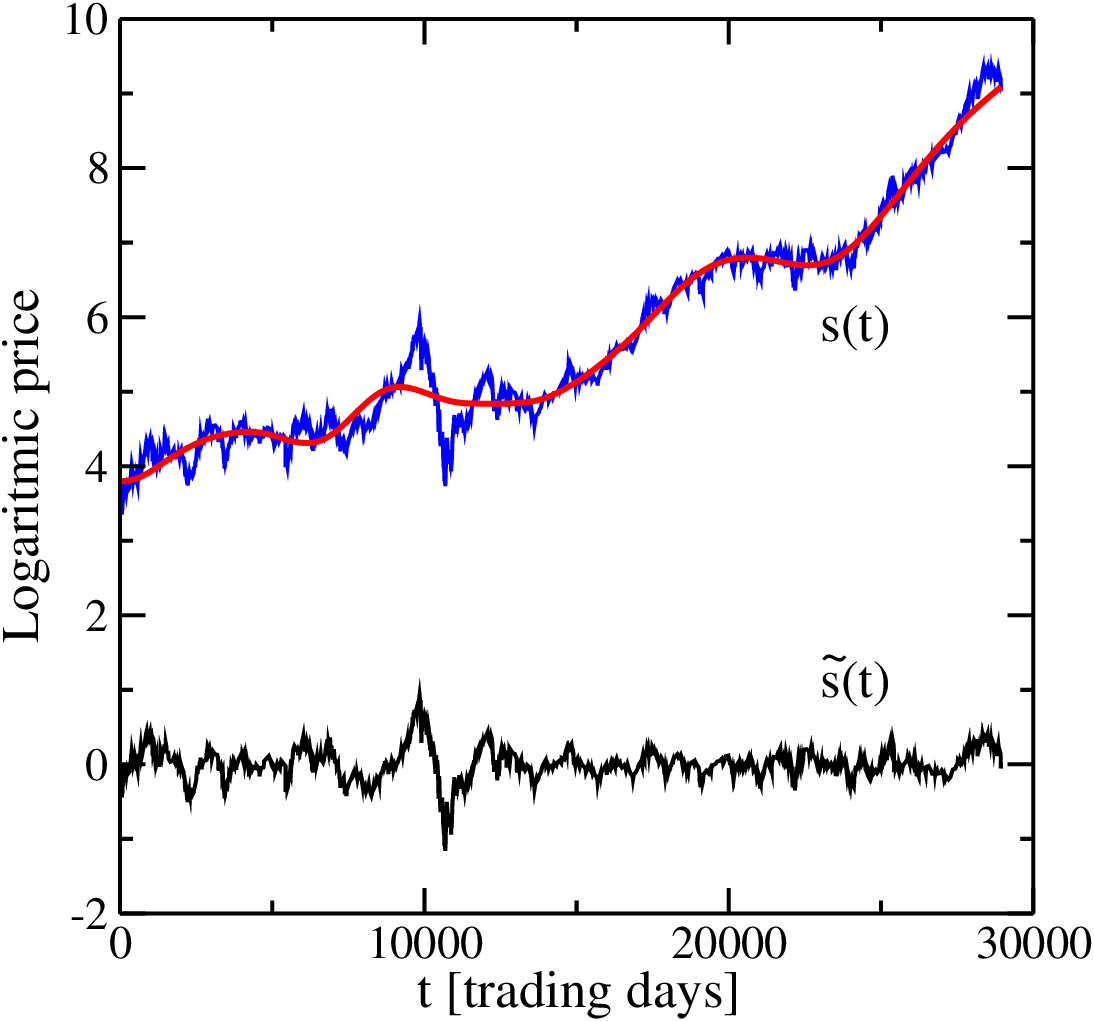}
        \caption{The historic daily logarithmic closure prices,
          $S(t)$, of the Dow Jones Industrial Average~(DJIA) over the
          period from May 26, 1896 to June 5, 2001. The upper curly
          curve is the raw logarithmic DJIA price $s(t)=\ln S(t)$,
          while the smooth curve represents the drift on a scale
          larger then $1000$ trading days. The lower curly curve
          represents the wavelet filtered logarithmic DJIA data,
          $\tilde{s}(t)$, defining the fluctuations of $s(t)$
          around the drift.}
        \label{Fig:DJIA}
    \end{center}
\end{figure}

%
%
It is well-known that many historic financial time series posses an
(often close to exponential) positive drift over long time scales. If
such a drift is present in the analyzed time series, one can obviously
not compare directly the histograms for positive and negative levels
of return. Since we in this paper mainly will be interested in making
such a comparison, one has to be able to reduce the effect of the
drift significantly. One possibility for detrending the data is to use
deflated asset prices.  However, in the present study we have chosen
an alternative strategy for drift removal based on the use of
wavelets~\cite{Book:NR-1992}, which has the advantages of being
non-parametric and does not rest on any economic theory whatsoever.
This technique has been described in detail
elsewhere~\cite{Simonsen-EPJB-2002}, and will therefore not be
repeated here. It suffices to say that this wavelet technique enables
a separation of the original time series into a short scale
(detrended) time series $\tilde{s}(t)$ and a drift term $d(t)$ so that
$s(t)=\tilde{s}(t)+d(t)$. In Fig.~\ref{Fig:DJIA}, we see the effect of
this procedure on the whole history of one of the major US economical
indicators, namely the Dow Jones Industrial Average~(DJIA). In this
particular example, which is the one used in the analysis, the
separation is set to $1000$ trading days,
corresponding to roughly $4$ calendar years.

Based on $\tilde{s}(t)$ for the DJIA, the empirical investment horizon
distributions, $p(\tau_\rho)$, can easily be calculated for various
levels of return $\rho$. In Fig.~\ref{Fig:DJIA_PDF} these empirical
distributions for $\rho=0.05$~(open circles) and $\rho=-0.05$~(open
squares) are presented. The solid lines in this figure are the maximum
likelihood fits of the empirical data to the functional
form~(\ref{fit-func}).  It is indeed observed that the generalized
Gamma distribution, Eq.~(\ref{fit-func}), fits the empirical data well
for both positive and negative levels of return. It has been checked
separately that the quality of the fits are of comparable quality for
other values of $\rho$. However, as $\left|\rho\right|$ becomes large,
the empirical distributions are hampered by low statistics that makes
the fitting procedure more difficult.
 
The most interesting feature that can be observed from
Fig.~\ref{Fig:DJIA_PDF}, is the apparent asymmetry between the
empirical investment horizon distributions for $\rho=\pm0.05$. In
particular, for $\rho=-0.05$ there is a higher probability, as
compared to what is observed for $\rho=0.05$, to find short investment
horizons, or in other words, {\em draw-downs} are faster then {\em
  draw-ups}. Consequently, one might say that there exists a gain-loss
asymmetry!  This result is in agreement with the drawdown/drawup
analysis presented in Ref.~\cite{outl2}.  Similar results to those
presented here have also been obtained for SP500 and NASDAQ.

\begin{figure}[t]
    \begin{center}
        \leavevmode
        \includegraphics*[width=0.8\columnwidth,height=0.43 \columnwidth]{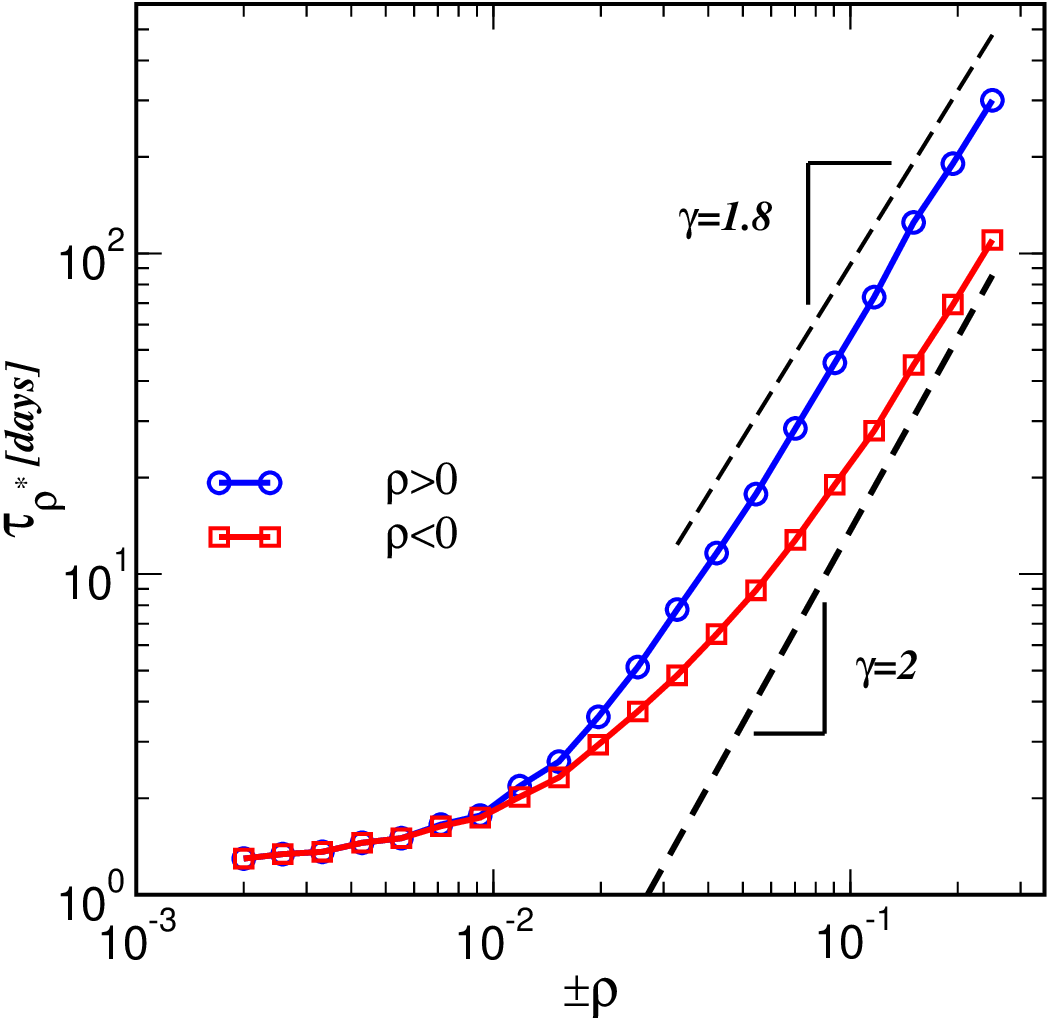}
        \caption{The optimal investment horizon $\tau^*_\rho$ 
          for positive~(open circles) and negative~(open squares)
          levels of return $\pm \rho$. In the case where $\rho<0$ one
          has used $-\rho$ on the abscissa for reasons of comparison.
          If a geometrical Brownian price process is assumed, one will
          have $\tau^*_\rho\sim \rho^\gamma$ with $\gamma=2$ for all
          values of $\rho$. Such a scaling behaviour is indicated by
          the lower dashed line in the graph.  Empirically one finds
          $\gamma\simeq 1.8$~(upper dashed line), only for large values of the return.}
        \label{Fig:DJIA_PDF_scaling}
    \end{center}
\end{figure}

Figure~\ref{Fig:DJIA_PDF_scaling} depicts the optimal investment horizon
{\it vs} level of return. From this figure it is observed that the
asymmetry feature found for a return level of $5\% $ is not unique.
For the smallest levels considered, $\left|\rho\right|\sim 10^{-3}$,
no asymmetry can be detected.  However, as $\left|\rho\right|$ is
gradually increased, the asymmetry starts to emerge at
$\left|\rho\right|\sim 10^{-2}$. By further increasing the level of
return, a state of saturation for the asymmetry appears to be reached.
In this state the asymmetry in the optimal investment horizon for the
DJIA is almost $200$~trading days.

These findings in fact confirms the saying in the financial industry
that {\em it takes time to drive up prices}. From this analysis, one
may add {\em compared to driving them down}, a result that coincides
with the common believe that the market reacts more violently to
negative information than to positive.
To our knowledge, this is the first time that such
statements have been founded in a quantitative analysis.  The
investment horizon distributions are, in fact, ideal tools for
addressing such questions quantitatively. Before arriving at the
conclusion of this paper, we will take the opportunity to speculate
about the reason for this asymmetry. Firstly, it cannot be due to any
residues of the drift left by the wavelet analysis. The economic
expansion in the 20'th century (corresponding to the time period of
the data) has in general been positive. Hence, if this was an effect
of the drift, it should be expected that the waiting times for
$\rho>0$ should be the shortest. However, from the data we find it to
be the other way around. Hence, we are led to conclude that this
asymmetry is reflecting the market dynamics.  Both from the point of
view of risk management and of market psychology, it makes sense that
market participants reacts faster to ``bad'' news than ``good'' news.

%
%
In conclusion, we have considered inverse statistics in economics. It
is argued that the natural candidate for such statistics is what we
call the investment horizon distribution. Such a distribution,
obtained from the historic data of a given market, indicates the time span
an investor historically has to wait in order to obtain a predefined
level of return. The distributions are parametrized excellently by a
shifted generalized Gamma distributions for which the first moment
does not exist. The typical waiting time, for a given level of return
$\rho$, can therefore be characterized by {\it e.g.} the time position
of the maximum of the distribution, {\it i.e.} by the {\em optimal}
investment horizon. By studying the behaviour of this quantity for
positive~(gain) and negative~(loss) levels of return, a very
interesting and pronounced gain-loss asymmetry emerges. It is concluded
that this asymmetry is part of the market dynamics.

\section*{Acknowledgments}

We would like to thank E. Aurell, S. Maslov and Y.-C. Zhang for usefull comments.


\end{document}